# Three Years of Mira Variable Photometry: What Has Been Learned?

*Dale E. Mais*
*Palomar Community College*
dmais@ligand.com

*David Richards*
*Aberdeen & District Astronomical Society*
david@richweb.f9.co.uk
&
*Robert E. Stencel*
*Dept. Physics & Astronomy*
*University of Denver*
rstencdel@du.edu

**Abstract**

The subject of micro-variability among Mira stars has received increased attention since DeLaverny et al. (1998) reported short-term brightness variations in 15 percent of the 250 Mira or Long Period Variable stars surveyed using the broadband 340 to 890 nm "Hp" filter on the HIPPARCOS satellite. The abrupt variations reported ranged 0.2 to 1.1 magnitudes, on time-scales between 2 to 100 hours, with a preponderance found nearer Mira minimum light phases. However, the HIPPARCOS sampling frequency was extremely sparse and required confirmation because of potentially important atmospheric dynamics and dust-formation physics that could be revealed. We report on Mira light curve sub-structure based on new CCD V and R band data, augmenting the known light curves of Hipparcos-selected long period variables [LPVs], and interpret same in terms of [1] interior structure, [2] atmospheric structure change, and/or [3] formation of circumstellar [CS] structure. We propose that the alleged micro-variability among Miras is largely undersampled, transient overtone pulsation structure in the light curves. 

## 1. Introduction

From European Space Agency's High Precision Parallax Collecting Satellite, HIPPARCOS (ESA, 1997) mission data, deLaverny et al. (1998) discovered a subset of variables (15 percent of the 250 Mira-type variables surveyed) that have exhibited abrupt short-term photometric fluctuations, within their long period cycle. All observations were made in a broadband mode, 340 to 890 nm, their so-called Hp magnitude. They reported variation in magnitude of 0.23 to 1.11 with durations of 2 hours up to almost 6 days, preferentially around minimum light phases. Instrumental causes could not be identified to produce this behavior. Most of these variations are below the level of precision possible with purely visual estimates of the sort collected by AAVSO, but may contribute to some of the scatter in visual light curves. 51 events in 39 M-type Miras were detected with HIPPARCOS, with no similar variations found for S and C-type Miras. These short-term variations were mostly detected when the star was fainter than Hp = 10 magnitude including one star at Hp = 13 magnitude and one at Hp = 8.3. For 27 of the original 39 observations, the star underwent a sudden increase in brightness. From their study, deLaverny et al. found that 85% of these short-term variations were occurring around the minimum of brightness and during the rise to the maximum, at phases ranging from 0.4 to 0.9. No correlation was found between these phases and the period of the Miras, but that brightness variations do occur preferentially at spectral types later than M6 and almost never for spectral types earlier than M4. Similar results were reported by Maffei & Tosti (1995) in a photographic study of long period variables in M16 and M17, where 28 variations of 0.5 mag or more on timescales of days were found among spectral types later than M6. Schaeffer (1991) collected reports on fourteen cases of flares on Mira type stars, with an amplitude of over half a magnitude, a rise time of minutes, and a duration of tens of minutes. In analogy to the R CrB phenomenon, brightness variation could also be consequence of dust formation (fading) and dissipation (brightening) in front of a star's visible hemisphere. Future narrow band infrared interferometric observations will help resolve this.

Recently, Wozniak, McGowen and Vestrand (2004) reported analysis of 105,425 I-band measurements of 485 Mira-type galactic bulge variables sampled every other day, on average, over nearly 3 years as a subset of the OGLE project. They failed to find any significant evidence for micro-variability, to a limit of 0.038 I-band events per star per year. They conclude that either Hipparcos data are instrumentally challenged, or that discovery is subject to metallicity or wavelength factors that minimize detection in I-band among galactic bulge objects. In contrast, Mighell and Roederer [2004] report flickering among red giant stars in the Ursa Minor dwarf spheroidal galaxy, including detection of low-amplitude variability in faint RGB stars on 10 minute timescales! However, Melikian [1999] provides a careful analysis of the light curves for 223 Miras based on Hipparcos data, finding that 82 stars [37%] show a post-minimum hump-shaped increase in brightness on the ascending branch of the light curve. Melikian advocates that differing physical processes and perhaps stellar properties – e.g. later spectral types, longer periods and higher luminosity - differentiate behaviors among these stars.

The purpose of this report is to provide V and R-band photometry of objects related to the deLaverny et al. results, with dense temporal sampling. We find a similar lack of micro-variability as noted by Wozniak et al., but do confirm facets of the Melikian report. This suggests that these phenomenon can be placed in a larger context of pulsational variations and episodic dust formation, with implications for ongoing spectroscopic and interferometric observations of mid-infrared studies of LPV stars.

## 2. Observations and Data Reduction

Our target list primarily was drawn from the objects listed by deLaverny et al. (1998), although limited to the northern sky. Of the 39 M type Mira's described therein, 20 are relatively bright and visible from the northern hemisphere. Because of the efficiency of automated sampling, we augmented this list with additional M type Miras and the brightest C, CS and S stars where one can obtain good signal to noise with low to moderate resolution spectroscopy on a small telescope. These stars and associated characteristics are detailed in Table 1.

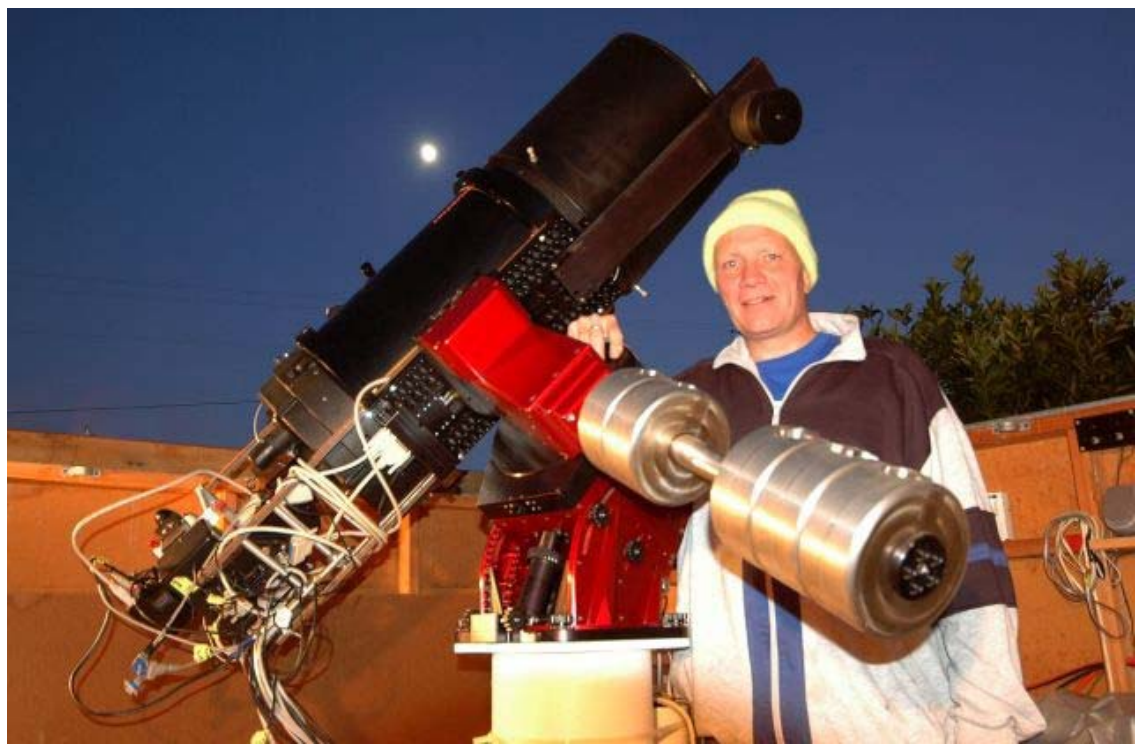

**Figure 1. Dale Mais and his photometry machine co-located near Valley Center, California.**

These along with a variety of brighter S and C type stars were also chosen. Brighter stars were chosen since they represented stars with magnitudes such that moderate resolution spectroscopy could be performed as part of the monitoring process. To accomplish this in a semi-automated manner, the telescope, camera and filter wheel are controlled by a single computer using Orchestrate software (www.bisque.com). Once the images are reduced, a script written by one of the authors (David Richards) examines the images performing an image link with TheSky software (www.bisque.com). The images obtained in this manner are stamped both with the name of the variable star, since this was how Orchestrate was instructed to find the object, and the position of the image in the sky. This allows TheSky to quickly perform the links with its USNO database. Once the astrometric solution is accomplished, the program reads through a reference file with the pertinent data such as reference star name and magnitudes along with variable star of interest. The input file is highly flexible, stars and filter magnitudes of reference stars can be added freely as image data require. This file only needs to be created once, especially convenient for a set of program stars, which will have continuous coverage over time. There is no need for entering magnitude information of reference stars in a repeated manner. The results file is readily imported to spreadsheet software, where the various stars and their magnitudes can be plotted, almost in real time. This is an important aspect of this project, the ability to see changes (flare-ups) quickly and as a result respond to these changes with spectroscopic observations.

(a) XZ Herculis

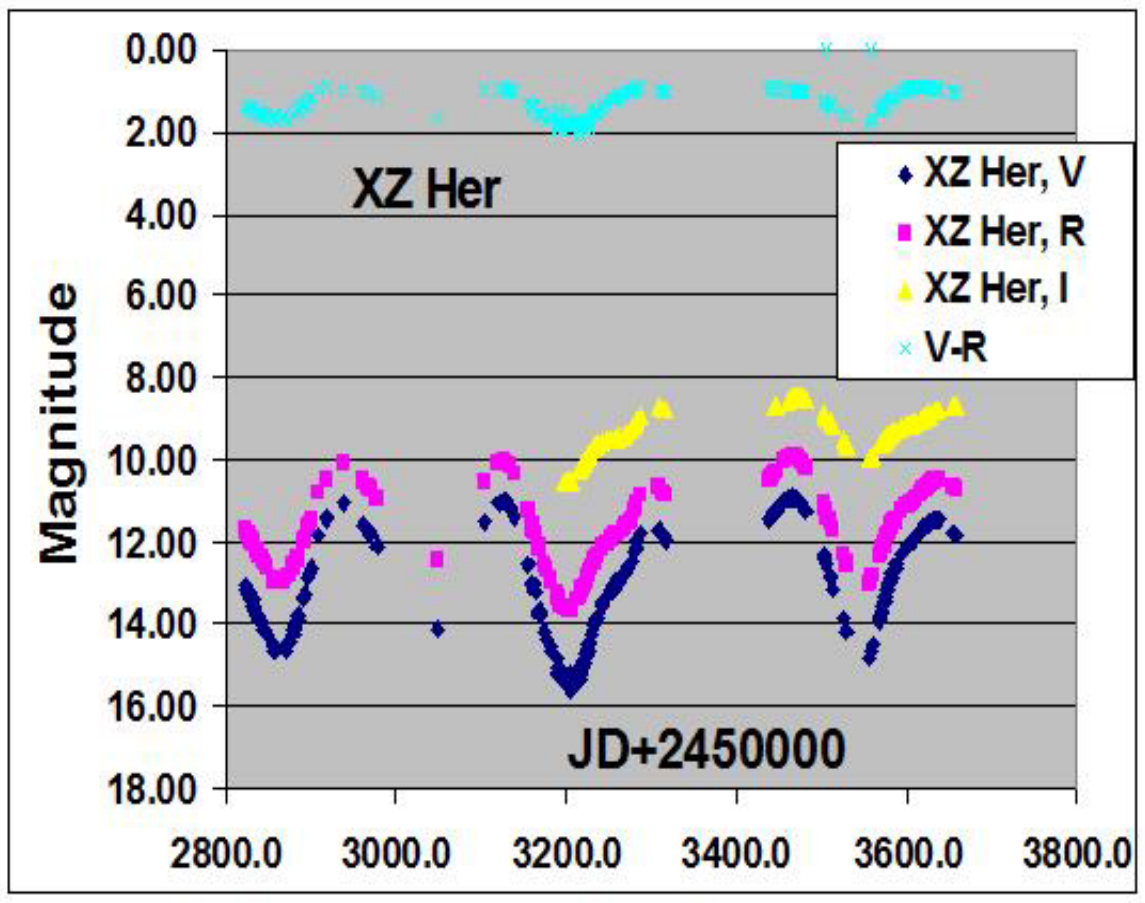

(b) S Serpentis

**Figures 2: Representative light curves. Additional examples can be found in Mais et al. 2005 SAS Conference Proceedings. These data will be submitted to the AAVSO archives.**

Photometry was conducted with an Astrophysics 5.1-inch f/6 refractor located in rural San Diego county, California, using an ST-10XME camera and 2x2 binned pixels and the Johnson V and R filters. Images were obtained in duplicate for each band and two reference stars used per variable star for analysis. Image reduction was carried out with CCDSOFT (www.bisque.com) and Source Extractor (Bertin and Arnout, 1996) image reduction groups and specially written scripts for magnitude determinations, which allowed for rapid, nearly real time magnitudes to be found (see below).

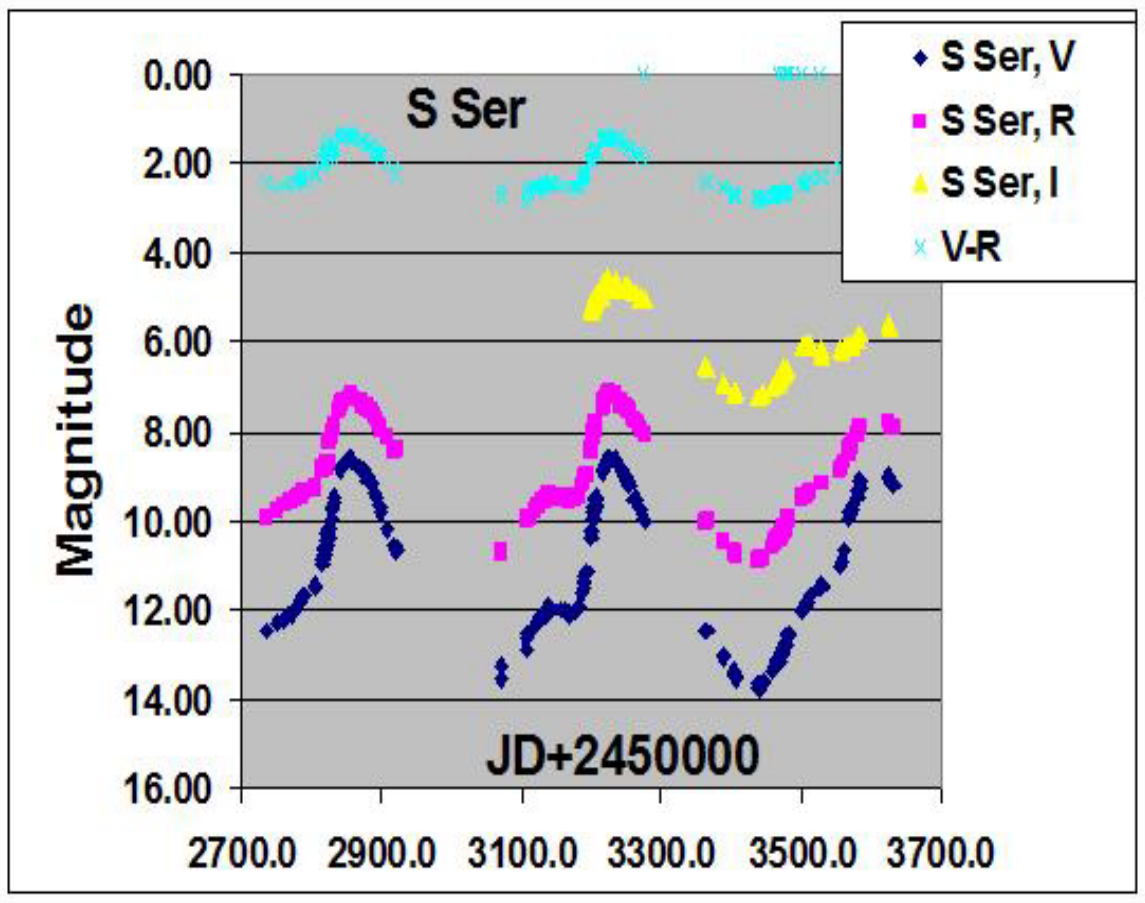

The project has been underway since 2003 and involves a total of 96 stars, 20 M type Miras, 19 S types and the remainder C types. While there are certainly many more of these type stars, only those that had a significant part of their light curve brighter than visual magnitude 8 were considered, due to magnitude limitations in the spectroscopy part of the project. Fortunately, these stars are much brighter in the R and I bands, often by 2-4 magnitudes when compared to their V magnitudes, and many of the interesting molecular features are found in this region of the spectrum. The photometric analysis involves using 2 different reference stars. Their constant nature is readily discerned over the time period by the horizontal slope of their light curves, both in the V and R bands. After considerable effort, magnitudes are now determined at the 0.02 magnitude level. Thus any flare-ups in the range of 0.1 magnitude and brighter should be readily discerned.

Early on it was felt that semi-automating the process was the best way to proceed. The use of a precision, computer-controlled mount (Paramount, **www.bisque.com**) along with the suite of software by Software Bisque got the project rapidly underway. TheSky in conjunction with CCDSOFT lends itself to scripting, and a script was put together that automated the magnitude determinations. To give an example of how this has streamlined the effort, on a typical night, initially using Orchestrate and later using a script developed by coauthor David Richards to control the telescope, camera and filter wheel, 40 stars, visible at the time, are imaged in duplicate in each of the V and R bands. This takes about 1 hour. Reduction of the images using image reduction groups in CCDSOFT takes another 5 minutes. The script that determines the magnitudes takes about 10 minutes to churn its way through all the images. Within another 20 minutes, the data, via spreadsheet, has been added to each variable stars growing light curve. Thus in less than 2 hours all of the program stars have been observed and their results tallied. Until more of program stars rotate into view, one is free to pursue spectroscopic examination of the program stars, establishing baseline observations. Another portion of this effort included standardizing the reference stars in each of the fields using the Landolt standards (Landolt, 1983). Once this is done all previous and subsequent observations of the variable stars will have their magnitudes expressed in absolute Johnson-Cousins magnitudes.

## 3. Results & Analysis

### 3a. Evidence for flares?

Three years of monitoring of 96 Mira-like variables, including all the northern objects included in the Hipparcos report [deLaverny et al. 1998] has yielded minimal evidence for flare-like changes in V or R band. Best cases include 0.2 and 0.3 magnitude increases near minimum light in RR Boo. This object was reported by Guenther and Henson [2001] to have shown a one-time 0.8 mag flare. Other marginal cases in our data include CE Lyr, X CrB and DH Lac. In the case of CE Lyr, the proximity of a faint stellar companion could contribute to jumps in automated photometry as the variable changes around minimum. Otherwise, most variables show smooth light variation with no hint of flare-like fluctuations at a level of 0.01 mag.

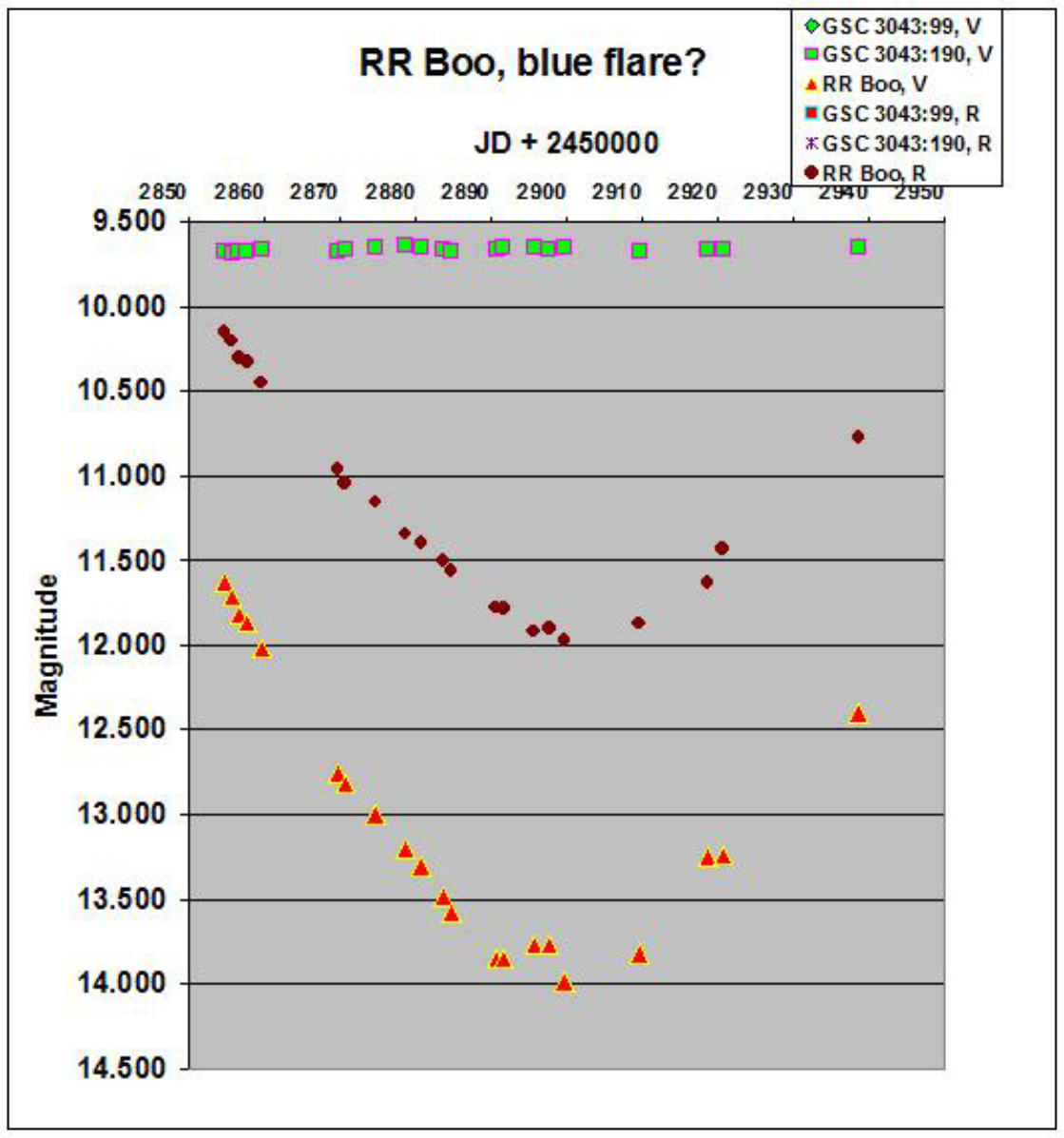

**Figure 3a. variation near minimum light, RR Boo**

### 3b. Light curve "bump" phenomena

In contrast to high-frequency events like flares [hours or days], examining the light curves for the stars observed between 2003 and 2006 revealed persistent low frequency changes on timescales of weeks. Following the discussion of these by Melikian [1999], we label these "bumps" in the Mira light curves. Good examples of this are seen in the light curves of RT Boo, R CMi, X CrB, U Cyg, XZ Her [Fig.2a], S Ser [Fig.2b], RU Her, U Cyg and R Lyn. Some of these are seen in visual light curves compiled by AAVSO and AEFOV, but others are seen at levels below the ~0.1 mag precision typical of visual observations. This is one of the important benefits of high precision photometry. Most light curve bumps are non-recurring and seem to appear after especially deep minimum light. The correlation of bumps with Mira properties deserves further attention. A few extremes in our sample are noted: double maxima in T Cam, S Cas, RR Her, S Cep, RS Cyg, RR Her and Y Per. Two stars show the bump feature post-maximum light: V CrB and T Dra.

### 3c. Period determination

Period finding was performed using PerAnSo software suite by Tonny Vanmunster, [http://users.skynet.be/fa079980/peranso/index.htm] which reports best fits using the ANOVA method. Initial application shows agreement with literature periods for most variables, within a few percent and with errors in fit of 1 to 20 days, depending on the data interval and phase coverage thus far. More work is in progress.

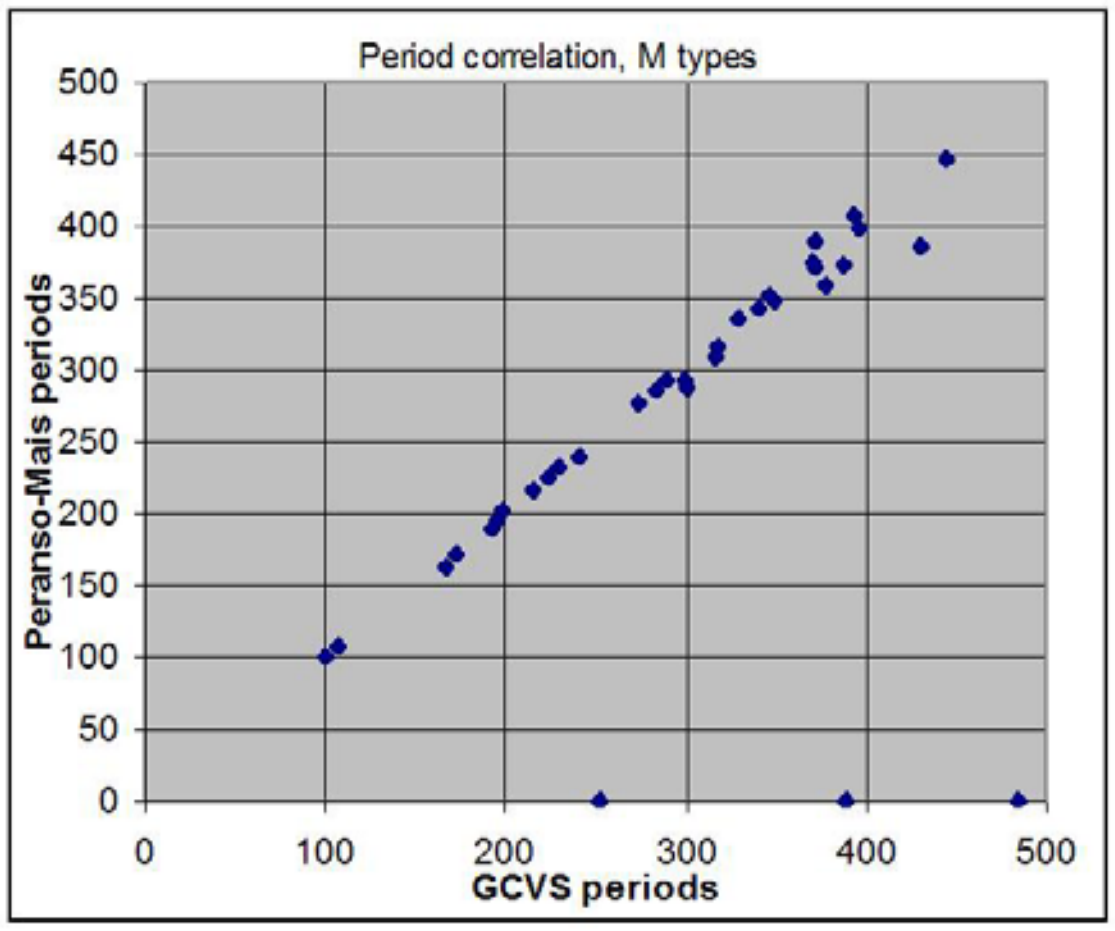

**Figure 3b: correlation between M-type Miras newly derived periods and older GCVS determinations.**

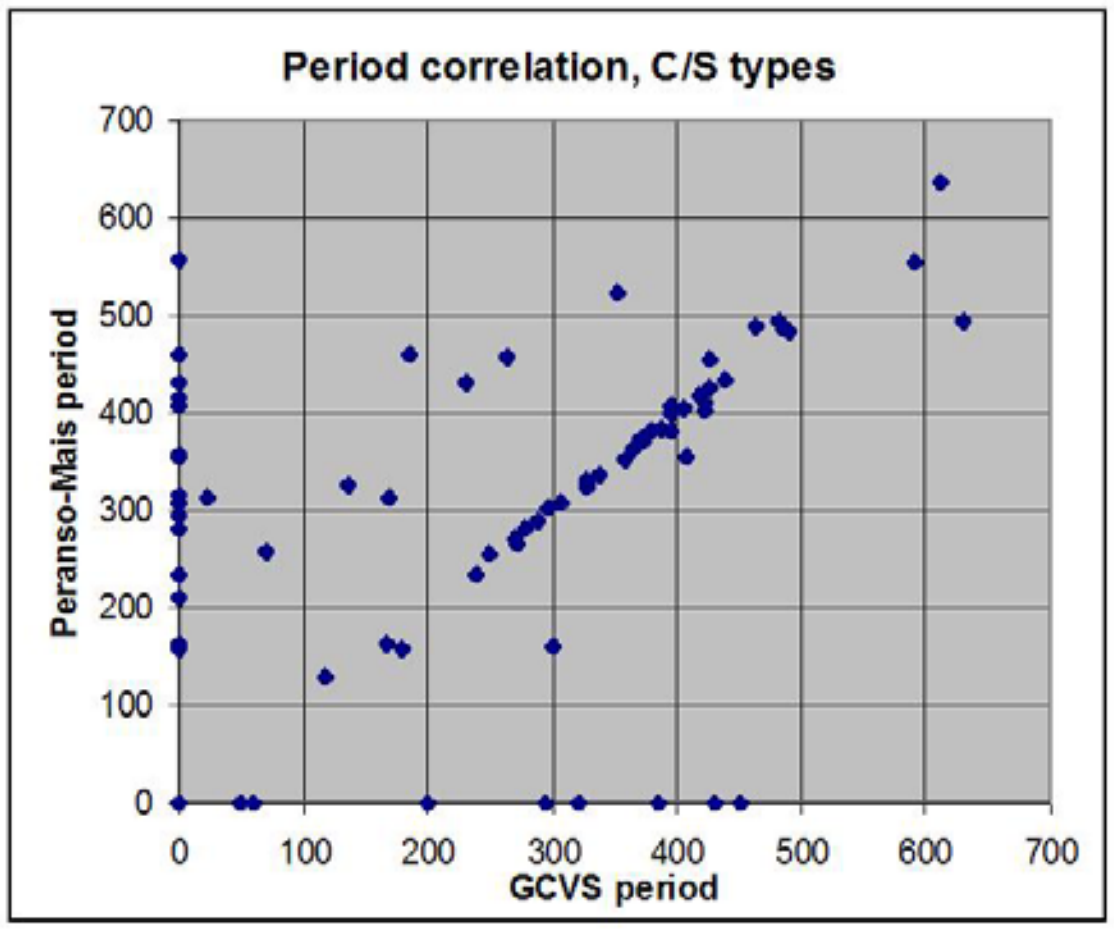

**Figure 3c: correlation between C&S-type Miras newly derived periods and older GCVS determinations.**

## 4. Conclusions

Among our conclusions based on V and R band measurements, with ~10 millimag precision, of nearly one hundred brighter Mira type stars are:

[1] flare events are rare, and statistically similar to the OGLE result for I band monitoring of 0.038 events per star per year, with some evidence that "flares" are bluer in color;

[2] we are confirming indications of correlations between depth of minima and occurrence of a "bump" or change of slope on the ascending branch of some light curves [cf. Melikian 1999];

[3] our coverage of approximately 3 cycles is sufficient to confirm the majority of previously published periods;

[4] we hypothesize that bump phasing and contrast varies with internal structure and opacity in analogy with similar phenomena among the "bump Cepheids" and deserves further study.

## 5. Acknowledgements

The authors wish to acknowledge the assistance of Thomas Bisque of Software Bisque for many useful discussions and help with the scripting, and of the AAVSO for its compilation of variable star visual light curves referenced in this paper, and the estate of William Herschel Womble for partial support of University of Denver astronomers participating in this effort.

.

.

.

.

.

## Tables

Table 1a: M type Mira Stars monitored

| Star* | Sp.Type, Vmax | Period, Epoch* GCVS IV | Period,Epoch* Anova/Peranso | Notes |
|---|---|---|---|---|
| SV And** | M5e, 7.7 | 316.21, 42887 | 308.64, 52892.3 | |
| R Boo | M3e, 6.2 | 223.40, 44518 | 224.22, 53465.3 | |
| RR Boo** | M2e, 8.3 | 194.70, 43047 | 194.74, 53197.7 | |
| RT Boo** | M6.5e, 8.3 | 273.86, 42722 | 276.12, 53225.5 | |
| W Cnc | M6.5e, 7.4 | 393.22, 43896 | 406.36, 53336.8 | |
| R CVn | M5.5e, 6.5 | 328.53, 43586 | 335.14, 53425.5 | |
| V CVn | M4e, 6.5 | 191.89, 43929 | 189.21, 53210.1 | |
| R Cas | M6e, 4.7 | 430.46, 44463 | 385.8, 53533.9 | dP/dt? |
| T Cas | M6e, 6.9 | 444.83, 44160 | 447.18, 53289.3 | |
| V Cas | M5e, 6.9 | 228.83, 44605 | 232.71, 53285.6 | |
| T Cep | M5.5e, 5.2 | 388.14, 44177 | 372.67, 53287.2 | |
| R Cet** | M4e, 7.2 | 166.24, 43768 | 163.30, 53221.6 | |
| X CrB** | M5e, 8.5 | 241.17, 43719 | 240.17, 52865.3 | |
| AM Cyg** | M6e, 11.3 | 370.6, 30075 | 374.47, 53345.2 | |
| T Eri** | M3e, 7.2 | 252.29, 42079 | --, -- | |
| RU Her** | M6e, 6.8 | 484.83, 44899 | --, -- | |
| SS Her** | M0e, 8.5 | 107.36, 45209 | 106.44, 53245.4 | |
| XZ Her** | M0, 10.5 | 171.69, 33887 | 171.15, 52939.3 | |
| R Hya | M6eTc, 3.5 | 388.87, 43596 | --, -- | |
| T Hya** | M3e, 6.7 | 298.7, 41975 | 292.44, 53492.1 | |
| X Hya** | M7e, 7.2 | 301.10, 41060 | 287.91, 53113.4 | dP/dt? |
| DH Lac** | M5e, 11.6 | 288.8, 41221 | 292.45, 52929.6 | |
| R Lmi | M6.5eTc:,6.3 | 372.19, 45094 | 390.02, -- | |
| W Lyr | M2e, 7.3 | 197.88, 45084 | 201.57, 53575.0 | |
| CE Lyr** | --, 11.7 | 318, 25772 | 315.90, 53259.0 | |
| HO Lyr** | M2e, 11.4 | 100.4, 30584 | 99.75, 53235.6 | |
| V Mon** | M5e, 6.0 | 340.5, 44972 | 342.51, 53312.7 | |
| RX Mon** | M6e, 9.6 | 345.7, 35800 | 351.62, 53192.9 | |
| Z Oph | K3ep, 7.6 | 348.7, 42238 | 347.52, 53450.6 | |
| R Peg | M6e, 6.9 | 378.1, 42444 | 358.17, 53350.0 | dP/dt? |
| RT Peg | M3e, 9.4 | 215.0, 45599 | 216.76, 53315.5 | |
| SW Peg | M4e, 8.0 | 396.3, 38750 | 398.43, 53482.9 | |
| S Per | M3Iab, 7.9 | 822, -- | 775.33, 53058.8 | |
| S Ser** | M5e, 7.0 | 371.84, 45433 | 371.69, 53234.1 | |
| AH Ser** | M2, 10.0 | 283.5, 36682 | 284.83, 53413.8 | |

*Stars are ordered by the traditional nomenclature: alphabetical by constellation, then by single letter R…Z, then double letters RR…ZZ, then AA…QQ and finally V###; MJD= JD – 2,400,000; GCVS: Gen.Catalog Var.Stars 4th ed., 1985 & http://www.sai.msu.su/groups/cluster/gcvs/gcvs/

** Hipparcos "flare stars" included in deLaverny et al. [1998]

.

.

.

.

Table 1b: C & S type Mira Stars monitored

| Star* | Sp.Type, V [max] | Period, Epoch* GCVS IV | Period, Epoch* Anova/Peranso | Notes |
|---|---|---|---|---|
| W And | S6,1e, 6.7 | 395.93, 43504 | 406.72, 53221.6 | |
| RR And | S6.5,2e, 8.4 | 328.15, 43390 | 327.17, 52917.8 | |
| ST And | C4.3e, 7.7 | 328.34, 38976 | 324.59, 53424.4 | |
| SU And | C6.4N8,8.0 | --, -- | 282.83, 53235.0 | dP/dt? |
| V Aql | C5.4N6,6.6 | 353, -- | 524.43, 53608.2 | dP/dt? |
| W Aql | S3.9e, 7.3 | 490.43, 39116 | 484.53, 53157.9 | |
| UV Aql | C5.4N4,11.1 | 385.5, 30906 | --, -- | |
| S Aur | C4-5N3,8.2 | 590.1, 42000 | 554.59, 53169.9 | dP/dt? |
| V Aur | C6.2eN3e,8.5 | 408.09, 43579 | 356.18, 53287.1 | dP/dt? |
| TX Aur | C5.4N3, 8.5 | --, -- | 163.30, 53370.8 | |
| EL Aur | C5.4N3,11.5 | --, -- | 234.60, 53064.8 | |
| FU Aur | C7.2N0,11.0 | --, -- | 157.04, 53265.2 | |
| R Cam | S2.8e, 6.97 | 270.22, 43978 | 270.53, 53153.5 | |
| S Cam | C7.3eR8,7.7 | 327.26, 43360 | 331.74, 53221.6 | |
| T Cam | S4.7e, 7.3 | 373.20, 43433 | 370.74, 53160.7 | |
| U Cam | C3.9N5, 11.0 | --, 43060 | --, -- | |
| RU Cam | C0.1K0, 8.1 | 22, -- | 313.48, -- | |
| ST Cam | C5.4N5, 9.2 | 300:, -- | 160.30, 53214.9 | |
| UV Cam | C5.3R8, 7.5 | 294., -- | --, -- | |
| W Cma | C6.3N, 6.35 | --, -- | 295.53, -- | |
| R Cmi | C7.1eJ, 7.25 | 337.78, 41323 | 335.77, 53114.8 | |
| T Cnc | C3.8R6, 7.6 | 482, -- | 495.05, 53396.9 | |
| V Cnc | S0e, 7.5 | 272.13, 43485 | 266.19, 53284.3 | |
| S Cas | S3.4e, 7.9 | 612.43, 43870 | 636.07, 53059.6 | |
| U Cas | S3.5e, 8.0 | 277.19, 44621 | 280.90, 52913.8 | |
| W Cas | C7.1e, 7.8 | 405.57, 44209 | 404.94, 53143.7 | |
| ST Cas | C4.4N3, 11.6 | --, -- | 408.53, 53307.0 | |
| V365 Cas | M5S7.2,10.2 | 136, -- | 325.73, 52844.8 | |
| S Cep | C7.4eN8,7.4 | 486.84, 43787 | 488.08, 52976.0 | |
| V CrB | C6.2eN2, 6.9 | 357.63, 43763 | 351.86, 53449.8 | |
| R Cyg | S2.5eTc, 6.1 | 426.45, 44595 | 425.24, 53550.6 | |
| U Cyg | C7.2eNp,5.9 | 463.24, 44558 | 488.61, 52919.2 | dP/dt? |
| V Cyg | C5.3eNp, 7.7 | 421.27, 44038 | 403.23, 53248.1 | |
| RS Cyg | C8.2eN0p,6.5 | 417.39, 38300 | 419.58, 53254.4 | |
| RV Cyg | C6.4eN5,10.8 | 263, -- | 458.72, 53592.6 | |
| RY Cyg | C4.8N, 8.5 | --, -- | 307.22, 53233.4 | |
| SV Cyg | C5.5N3, 11.7 | --, -- | 415.47, 53164.7 | |
| TT Cyg | C5.4eN3,10.2 | 118, -- | 129.03, 53255.9 | |
| YY Cyg | C6.0evN,12.1 | 388, 298261 | 384.62, 52941.8 | |
| AW Cyg | C4.5N3,11.0 | --, -- | 557.88, 52875.9 | |
| AX Cyg | C4.5N6, 7.85 | --, -- | 461.75, 52845.1 | |
| V460 Cyg | C6.4N1, 5.6 | 180:, -- | 158.31, 53183.8 | |
| T Dra | C6.2eN0, 7.2 | 421.6, 43957 | 410.17, 53283.2 | |
| UX Dra | C7.3N0, 5.9 | 168, -- | 314.27, 53246.2 | |
| R Gem | S2.9eTc, 6.0 | 369.91, 43325 | 374.13, 53359.8 | |
| T Gem | S1.5e, 8.0 | 287.79, 44710 | 289.15, 53153.5 | |
| TU Gem | C6.4N3, 9.4 | 230, -- | 431.03, 53100.2 | |
| NQ Gem | C6.2R9ev,7.4 | 70:, -- | 256.72, -- | |
| S Her | M4Se, 6.4 | 307.28, 45054 | 307.33, 53638.5 | |
| RR Her | C5.7eN0, 8.8 | 239.7, -- | 233.47, 53126.3 | |
| R Lep | C7.6eN6, 5.5 | 427.01, 42506 | 456.19, 53009.3 | |

| | | | |
|---|---|---|---|
| SZ Lep | C7.3R8, 7.4 | --, -- | --, -- |
| R Lyn | S2.5e, 7.2 | 378.75, 45175 | 382.52, 53124.2 |
| T Lyr | C6.5R6, 7.8 | --, -- | 430.79, 52892.2 |
| U Lyr | C4.5eN0, 8.3 | 451.72, 42492 | --, -- |
| HK Lyr | C6.4N4, 7.8 | --, -- | 354.11, 53182.0 |
| V614 Mon | C4.5JR5, 7.0 | 60:, -- | --, -- |
| V Oph | C5,2N3e,7.3 | 297.21, 45071 | 302.58, 52826.6 |
| TW Oph | C5.5N, 11.6 | 185:, -- | 459.35, -- |
| RT Ori | C6.4Nb, 9.7 | 321, -- | --, -- |
| BL Ori | C6.3NbTc,7.9 | --, -- | 316.01, 53036.8 |
| RX Peg | C4.4JN3, 9.7 | 629:, -- | 495.24, 52987.9 |
| RZ Peg | C9.1eNTc,7.6 | 438.7, 54248 | 433.93, 53096.8 |
| HR Peg | S5.1M4, 6.1 | 50:, -- | --, -- |
| Y Per | C4.3eR4, 8.1 | 248.6, 45245 | 255.81, 53083.3 |
| V466 Per | C5.5N5, 10.9 | --, -- | 358.8, 53288.8 |
| T Sgr | S4.5e, 7.1 | 394.66, 44897 | 382.65, 53049.7 |
| ST Sgr | S4.3e, 7.2 | 395.12, 40463 | 401.07, 53046.6 |
| AQ Sgr | C7.4N3, 9.1 | 199.6, -- | --, -- |
| V1942 Sgr | C6.4N2R8,6.7 | --, -- | 209.68, 52879.6 |
| FO Ser | C4.5R6, 8.5 | --, -- | --, -- |
| TT Tau | C4.2N3, 10.2 | 166.5, -- | 163.30, 53010.9 |
| SS Vir | C6.3eN, 6.0 | 364.14, 45361 | 361.85, 53231.7 |
| BD Vul | C6-7Ne, 9.3 | 430, 25758 | --, -- |

----------------------------------------------------------------------------------------------------

*Stars are ordered by the traditional nomenclature: alphabetical by constellation, then by single letter R…Z, then double letters RR…ZZ, then AA…QQ and finally V###; MJD= JD – 2,400,000; GCVS: Gen.Catalog Var.Stars 4th ed., 1985 & http://www.sai.msu.su/groups/cluster/gcvs/gcvs/

-o-